\documentstyle{mn}
\input{epsf}
\begin{document}
\def\p0437{PSR J0437$-$4715}

\title[Radio Profile Instability in \p0437 due to Spiky Emission]
{Radio Profile Instability in the Millisecond Pulsar \p0437 due to Spiky Emission}

\author[M. Vivekanand]{M. Vivekanand \thanks{vivek@ncra.tifr.res.in} \\
National Center for Radio Astrophysics, TIFR,  Pune University Campus, P. O. Box 3, 
Ganeshkhind, Pune 411007, India.}

\date{}

\maketitle
\begin{abstract}
The instability in one of the components of the radio integrated profile (IP) of 
the millisecond pulsar \p0437 is correlated with the occurrence of the spiky 
emission in that component. It is speculated that the instability in other
components is also due to the spiky emission. This might have important 
implications for interpreting the individual components of the IP as independent 
beams or regions of emission.

\end{abstract}
\section{Introduction}

The coherent radio emission mechanism of rotation powered pulsars is as yet
an unsolved problem. The broad similarity in the coherent radio emissions 
from normal and millisecond pulsars (McConnell et al. \cite{MABE}, Gil \& 
Krawczyk \cite{GK}, Jenet et al. \cite{Jenet}, Vivekanad et al. \cite{VAM}, 
Kramer et al. \cite{Kramerb}, Vivekanand \cite{MV2000}) is intriguing; that 
a similar mechanism operates over three to four orders of magnitude 
difference in their periods and period derivatives is probably an important 
clue to the coherent radio emission mechanism of rotation powered pulsars.

Of equal importance are the differences in the radio emission from these two 
classes of pulsars. One of them is the {\bf slow} variation in the IP of 
millisecond pulsars (Camilo \cite{Camilo}, Backer \& Sallmen \cite{BS}, 
Vivekanad et al. \cite{VAM}, Kramer et al. \cite{Kramera}), over time scales 
ranging from several hundred periods to hours and even days, which is quite 
unlike the {\bf fast} mode changing phenomenon observed in some normal 
pulsars (Backer \cite{Backer}, Helfand et al. \cite{HMT}, Manchester \& Taylor 
\cite{MT}, Rathnasree \& Rankin \cite{RR}). Neither phenomenon has a 
satisfactory explanation so far.

Here it is shown that the slow IP variations seen in one of the components 
of \p0437 (Vivekanand et al. \cite{VAM}) are correlated with variations of 
the spiky emission in that component (Vivekanand \cite{MV2000}).

\section{Profile Instability in \p0437}

Figure ~\ref{fig1} shows IP of \p0437 at five different epochs, 
observed at 326.5 MHz using Ooty Radio Telescope (ORT); only a single linear 
polarization was available. From top to bottom the five panels correspond to (1) 
9 files observed from UT 15:27:59 to UT 15:42:08 on 7 Mar 1995, (2) 2 files 
observed at UT 15:43:57 and UT 15:45:49 on 7 Mar 1995, (3) 12 files observed 
from UT 16:04:17 to UT 16:23:22 on 7 Mar 1995, (4) 13 files observed from UT 
15:39:25 to UT 16:04:21 on 9 Mar 1995, and (5) 3 files observed at UT 15:51:21, 
UT 15:53:02 and UT 15:54:40 on 14 Mar 1995. Each file consists of 12\,125 useful 
periods of data. The instrument, the method of observation and data analysis are 
described in Ables et al. \cite{AMDV}, Vivekanand et al. \cite{VAM}, Vivekanand 
\cite{MV2000}, and Vivekanand \cite{MV2001}.

In fig.~\ref{fig1} the first panel from top corresponds to the profile labelled 
type B in fig.~8 of section 5 of Vivekanand et al. \cite{VAM}. The second panel 
contains the two files that show a transition from type B to type A, and the 
third and fourth panels together correspond to the profile labelled type A. The 
main difference between the profiles of fig.~\ref{fig1} here and fig.~8 of 
Vivekanand et al. \cite{VAM} is the more accurate period, differing by a few 
nanoseconds, that is used for folding the data (Vivekanand \cite{MV2001}).

On 7 Mar 1995, the peak flux of the third component of the IP of \p0437 
decreased from 3.3 to 2.6 over a period of $\approx$ 36 minutes (panels 1 to 3 
of fig.~\ref{fig1}). During this time the peak flux of its central component 
increased from 5.2 to 5.6, and that of its first component increased 
marginally from 2.1 to 2.3. These correspond to fractional flux variations of 
21.2\%, 7.7\% and 9.5\% respectively, which are much larger than the $\approx$ 
1\% to 2\% maximum allowed by purely random noise. Two days later \p0437 had 
an IP similar to that observed earlier (panel 4). Five days later its first 
component became stronger than its third component (panel 5), exactly contrary 
to its behavior a week earlier (panel 1).

Vivekanand et al. \cite{VAM} argued that these profile variations are intrinsic 
to \p0437, and are not an artifact due to the single polarization observations 
of ORT coupled with Faraday rotation in the variable ionosphere. In the next 
section it will be shown that the IP variations of component 3 are highly 
correlated with the occurrence of the spiky emission in that component.

\section{Probability of Occurrence of Spikes}

On account of limited signal-to-noise ratio of the data (Vivekanand \cite{MV2001}), 
only the 500 strongest spikes in each data file were analyzed; ideally one would 
have liked to completely separate the spiky from the non-spiky emission for this 
exercise. 

First, the 500 strongest spikes in each data file were identified and their 
positions (phases) were obtained as described in Vivekanand \cite{MV2001}. Next, 
these positions were plotted against the numbers of the periods in which they 
occurred, for those spikes which satisfied following two selection criterion: (1) 
their peak signal to noise ratio should be greater than 12.5, and (2) the error
on their position should be less than 20 micro seconds ($\mu$s). However the final 
result is not very sensitive to these two selection criterion.

Figure~\ref{fig2} plots the positions of 375 of the 500 strongest spikes in the 
data file labelled 50661527, as a function of the period number. This belongs to 
the profile category labelled B by Vivekanand et al. \cite{VAM}; the rms errors 
on the positions are less than the width of the dots. There are 52, 314 and 9 
spikes in the three main components of the IP, respectively. If the two selection 
criterion are dropped, one obtains 66, 421 and 13 spikes in the corresponding 
components of the IP, which are entirely consistent with earlier numbers. In file 
labelled 50661538 belonging to the same category, that was observed at UT 15:38:41 
on 7 Mar 1995, \p0437 was at one of its lowest average fluxes. The number of 
spikes in the three main components of the IP, with and without the selection 
criterion, are 39, 301 and 1, and 56, 442 and 2, respectively, which are again 
fractionally similar. Therefore the two selection criterion do not significantly 
alter the results of this article. 

Figure~\ref{fig3} plots the positions of the 500 strongest spikes in the data file 
labelled 50661612, which belongs to the category labelled A by Vivekanand et al. 
\cite{VAM}. Here the number of spikes in the three main components of the IP are 
46, 453 and 1, respectively; no spikes have been selected out due to the high 
average pulsar flux in this data file. Clearly the fraction of spikes in the third 
component of the IP has decreased significantly compared to category B. This is
verified in a visual check of as many periods as is reasonably possible. Even in
file labelled 50661538, several periods had spikes in the third component of the 
IP, that were not part of the 500 strongest spikes.

To test this more rigorously, the above numbers were totalled for all files
in each category of the IP; the results are shown in Table 1. The 9 files of 
panel 1 in fig.~\ref{fig1} yielded totally N0 = 3598 spikes; the rest of the 
4500 spikes were ignored due to the two criterion mentioned above. Out of 
these, N1 = 408 spikes occurred in the first component of the IP, N2 = 3155 
spikes occurred in the second component, and the N3 = 35 spikes occurred in 
the third component. This was for the type B profile. N0 is 12\,405 for the 
25 files of type A profile (panels 3 and 4 in fig.~\ref{fig1}). From this 
one would have expected N1 = 408 / 3598 $\times$ 12\,405 $\approx$ 1407 $\pm$ 
38, whereas one obtained the smaller number 1255. Similarly one expected N2 
= 3155 / 3598 $\times$ 12\,405 $\approx$ 10\,878 $\pm$ 104 in the second 
component, but obtained the larger number 11\,144. But the most interesting 
number is the expected N3 = 35 / 3598 $\times$ 12\,405 $\approx$ 121 $\pm$ 11, 
which is significantly larger than the observed value of 6; and this occurs 
in spite of the average pulsar flux increasing by about 14\% (Table 1). 
Clearly the decrease in rate of occurrence of spikes in the third component
of the IP of \p0437 is correlated with the decrease in average flux of that 
component. 

This trend continues in the last row of table 1 (panel 5 of fig.~\ref{fig1}); 
but caution is required here. Since one selects the 500 strongest spikes in 
each data file, the results should not depend upon the average flux of \p0437 
to the zeroth order. But for very low average flux of \p0437 one expects to 
see no spike in the third component of the IP, since the probability of 
occurrence there is very low anyway (Vivekanand \cite{MV2000}). This is 
probably what is happening in the last row of table 1; the average number of 
spikes per file is significantly lower here.

\begin{table}
\begin{tabular}{ccccccc}
\hline
\hline
PANEL & FILES & N0 & N1 & N2 & N3 & $\left < FLUX \right >$ \\
\hline
\hline
1      & 9  & 3598    & 408    & 3155    & 35 & 3.10 $\pm$ 0.09 \\
3 \& 4 & 25 & 12\,405 & 1\,255 & 11\,144 & 6  & 3.54 $\pm$ 0.06 \\
5      & 3  & 344     & 33     & 311     & 0  & 2.09 $\pm$ 0.12 \\
\hline
\end{tabular}
\caption{
	 Statistics of occurrence of strong spikes in fig.~\ref{fig1}, 
	 for panel 1 (top row), panels 3 and 4 (middle row) and panel 
	 5 (bottom row). The second column contains the number of data 
	 files in each row. N0 is the total number of spikes in those 
	 files; N1, N2 and N3 are the number of spikes in the three main 
	 components of the IP of that row, respectively;
	 the boundaries of the components are shown by the dashed lines 
	 in fig.~\ref{fig2} and fig.~\ref{fig3}. The last column 
	 contains the average flux (in arbitrary units) of \p0437 in 
	 those panels, along with its rms error.
	}
\end{table}

The expected values of N1, N2 and N3 in the last panel of fig.~\ref{fig1}, based on
the type B profile, are 39 $\pm$ 6, 302 $\pm$ 17 and 3 $\pm$ 2; these are consistent 
with the observed values. One would have liked to see a larger value for N1 here, to 
be consistent with the increased flux of the first component of the IP; but one is 
probably dealing with small numbers due to limited data, as well as a significantly 
weaker pulsar flux in those files.

Table 1 has also been derived after removing the two selection criterion, and the
results remain similar.

The N3 values in the 9 files of row 1 of table 1 range from 9 to 1. The first three 
files observed chronologically, labelled 50661527, 50661529 and 50661531, have N3 = 
9, 5 and 7, respectively. Now \p0437 was strong in these files and weakened 
systematically with time. Not surprisingly, N3 vales in the next six files are 3, 2, 
4, 1, 2 and 2; these files are labelled 50661533 through 50661542. It is clear that 
there was no hope of seeing any spikes in the third component of the IP in the next 
two files, viz. 50661543 and 50661545 in panel 2 of fig.~\ref{fig1}, even if they 
existed, due to the low average flux of 2.54 $\pm$ 0.04 in these two data files; 
which is why the single pulse mode of observation was discontinued after that time. 
However, after about 18 minutes, the average flux of \p0437 became high again; which 
is why the single pulse mode of observation was re-started once again. But N3 for 
the next 12 data files, belonging to panel 3 of fig.~\ref{fig1}, was a mere 5; only 
five data files had N3 = 1, and the rest had N3 = 0. More convincingly, N3 was a mere 
1 for the 13 data files belonging to panel 4 of fig.~\ref{fig1}; and these contain 
some of the highest average flux data on \p0437.

It is therefore concluded that the decrease in N3 between the first two 
rows of table 1 is a feature intrinsic to \p0437. That this decrease 
correlates with the decrease in the flux of the third component of the
IP proves the basic hypothesis of this article. The 
clinching evidence would have been if one had actually seen the decrease
in the number of spikes as time progresses (in figures similar to 
fig.~\ref{fig2} and fig.~\ref{fig3}), in the two data files labelled
50661543 and 50661545 of panel 2 of fig.~\ref{fig1}, which are believed 
to be in transition between categories B and A. Unfortunately, this is 
not possible due to the low average flux of \p0437 in these two data 
files.

\section{Summary and Discussion}

Vivekanand et al. \cite{VAM} showed that the IP of the milli second pulsar 
\p0437 changes significantly in shape over time scales of minutes and hours 
and even days. This is unlike the much faster mode changing known in normal 
pulsars. They argued that this behavior was intrinsic to \p0437, and not an 
instrumental effect. Here it is shown that profile variations in the third 
component of the IP of \p0437 are correlated with the rate of occurrence of 
the spiky emission in that component. Each period that contained any of the 
35 + 6 = 41 spikes, in the third component of the IP in Table 1, was visually 
checked for confirmation.

\begin{figure}
\epsfxsize=9.0cm \epsfbox{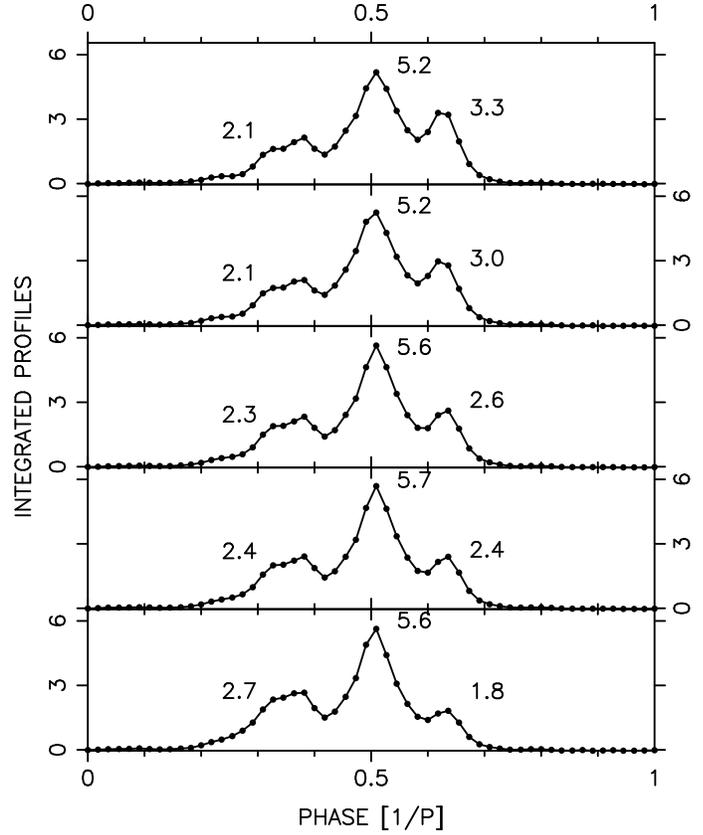}
\caption{
	 Integrated profiles of \p0437 observed at 326.5 MHz using ORT, at 
	 five different epochs during 7 to 14 Mar 1995. The abscissa is in 
	 units of the pulsar period (also known as phase within the period), 
	 which is changing due to its binary nature but is $\approx$ 5.757 
	 ms; the 56 time bins along the abscissa have a width each of 
	 $\approx$ 102.8 $\mu$s. The ordinates are in arbitrary units. The
	 area under each IP is normalized to a constant value. The peak 
	 values in the three principal components of the IP are listed for 
	 comparison.
	}
\label{fig1}
\end{figure}

\begin{figure}
\epsfxsize=9.0cm \epsfbox{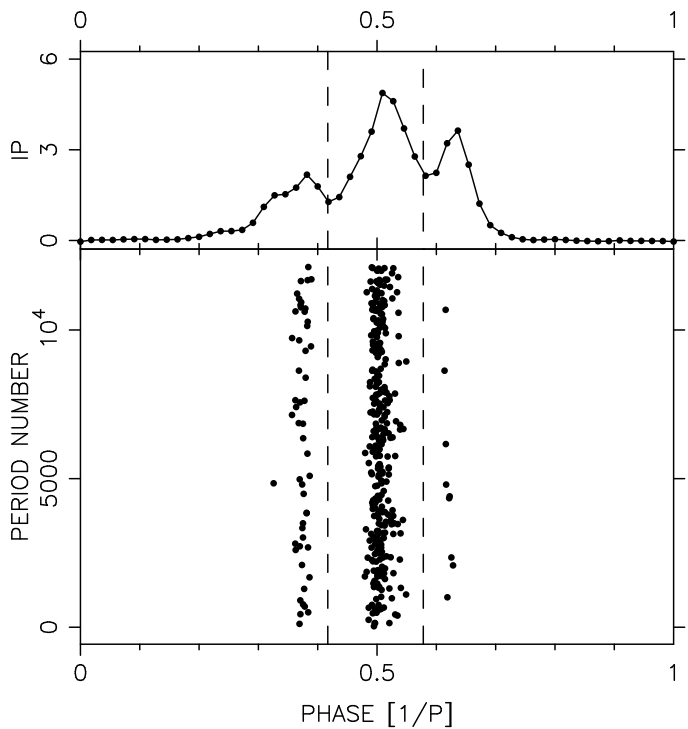}
\caption{
	 {\bf Top panel}: Integrated profile of 12\,125 periods in data file 
	 labelled 50661527, observed at UT 15:27:59 on 7 Mar 1995. This is one
	 of the nine files in the first panel of fig.~\ref{fig1}; abscissa and
	 ordinate are as in that figure. {\bf Bottom panel}: Ordinate is the 
	 period number. The dots represent the 375 largest spikes in the data 
	 file. The dashed vertical lines are drawn at abscissa 0.417 and 0.578, 
	 which are taken to be the practical boundaries between the three main 
	 components of the IP of \p0437.
	}
\label{fig2}
\end{figure}

\begin{figure}
\epsfxsize=9.0cm \epsfbox{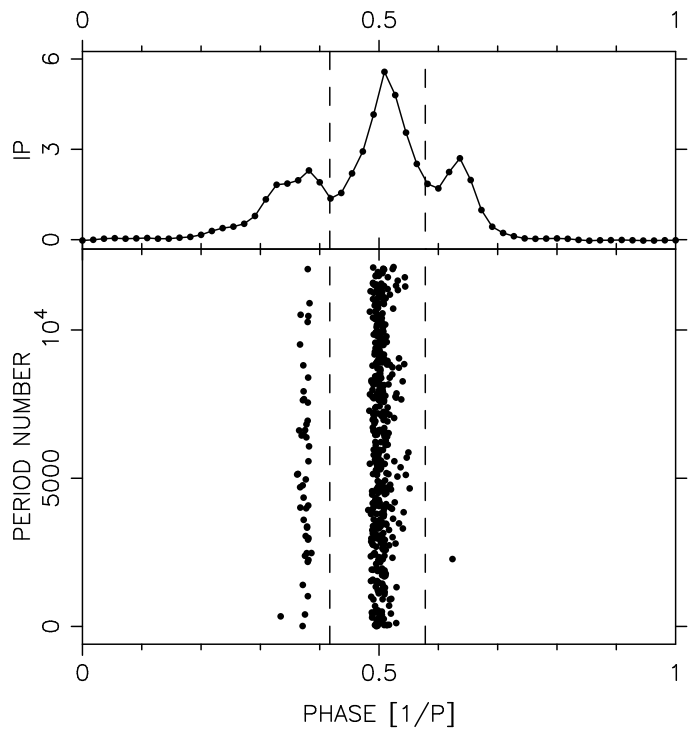}
\caption{
	 Same as fig.~\ref{fig2} above, but for the data file labelled 5066162, 
	 observed at UT 16:12:47 on 7 Mar 1995. This is one of the twelve files
	 in the third panel of fig.~\ref{fig1}.
	}
\label{fig3}
\end{figure}

Before proceeding further, two comments are in order:

\begin{enumerate}

\item The first concerns the availability of only a single polarization at
ORT. Navarro et al. \cite{NMS1997} show that the percentage of linear 
polarization of \p0437 at 438 MHz is highest at the peaks of components
2 and 3 in fig.~\ref{fig1} (also their fig.~1); component 1 has very low 
linear polarization. From their fig.~1 the position angle of linear 
polarization differs by less than $\approx 40^\circ$ at these two peaks. 
Therefore, it is possible that, in principle, differential Faraday rotation 
in a variable ionosphere at ORT can reduce the effective flux density of 
spikes in component 3 by $\approx 1 - \cos^2 ( 40^\circ ) \approx $ 41\% 
or less, in spite of the arguments of Vivekanand et al. \cite{VAM}. This 
issue can be resolved satisfactorily only when full polarization 
observations of \p0437 are available at 326.5 MHz. However, one would have 
expected an as frequent reduction of number of spikes in component 2 also,
with respect to the number of spikes in component 3; 
this is not observed, although the statistical basis of such a conclusion 
is not strong. Furthermore, it is quite possible that the mean position of 
spikes, in components 2 and 3 of fig.~\ref{fig1}, is not coincident exactly 
with the corresponding position of the peak of linear polarization. In such 
an eventuality, the effect discussed above would be less serious, since the 
linear polarization profiles in components 2 and 3 are fairly sharply peaked. 
Finally, similar profile variations have been noted by Bell et al. 
\cite{BBM1997} in \p0437 at 430 MHz in {\bf full polarization} observations. 
This author feels they might have wrongly attributed the observed profile 
variations to ``relative gain variation in the two polarization channels and 
the high degree of polarization of the emission observed across the profile''.

\item Vivekanand \cite{MV2000} has shown that the position of occurrence of
spikes is uncorrelated with their height; no simple relation exists between
these positions and the integrated profile (figures 5 and 6 of that paper).
Therefore this author can not think of any obvious selection effect involving 
these parameters that can produce the result of this paper.

\end{enumerate}

One would have liked to find a quantitative correlation between the rate of 
occurrence of spikes (Table 1) and the average flux in the other two components 
of the IP (fig.~\ref{fig1}) also. Unfortunately this is complicated by two 
factors.  Firstly, the average pulsar flux in a data file could be so low 
that the observed rate of occurrence of spikes in any component may be biased 
to a lower value. Secondly, the average flux of the spiky emission can also 
change in principle, from file to file and from component to component. To 
illustrate the problem, the fraction of N3 spikes in Table 1 decreases down 
the rows, consistent with decrease in flux of the third component of the IP in 
fig.~\ref{fig1}. However, the fraction of N1 spikes in Table 1 also decreases, 
although marginally; this is inconsistent with the increase in flux of the 
first component of the IP; this can easily arise due to the reasons mentioned 
above, as also due to a small change in the occurrence rate of spikes in the 
second component of the IP, which has about 90\% of the spikes anyway.

\p0437 has also been studied at $\approx$ 1400 MHz by several workers, but
this kind of behavior has not been reported so far. One possible reason is 
the low signal to noise ratio of their data. The other possible reason could 
be that \p0437 does not show such behavior at other frequencies. There is a 
precedence for this -- PSR B1821-24 shows radio profile instability at
1400 MHz but not at 800 MHz (Backer \& Sallmen \cite{BS}).

It is likely that the IP variations of the first component of \p0437, as well 
as any other component, are also due to the rate of spiky emission in that 
component, although that has not been proved rigorously here.

\begin{figure}
\epsfxsize=9.0cm \epsfbox{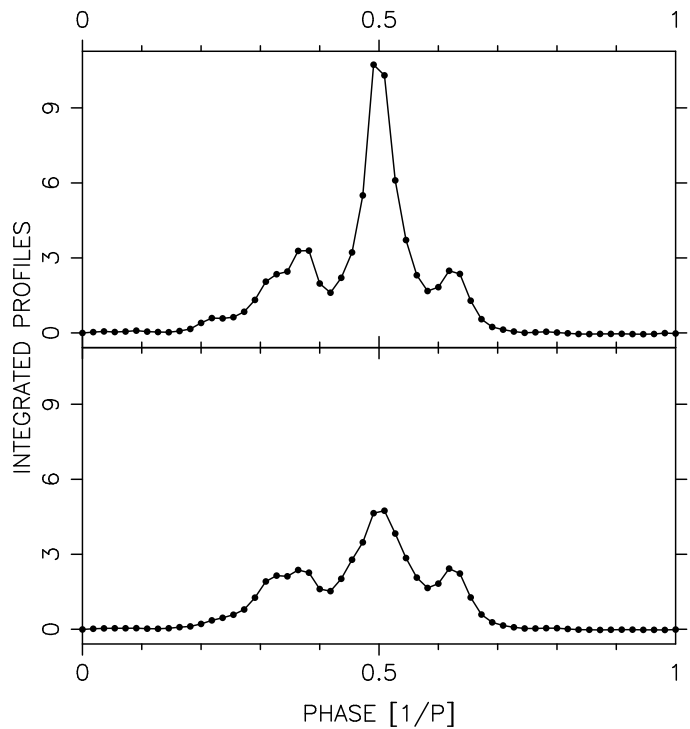}
\caption{
	 Integrated profiles of \p0437 from data in file labelled 50681546,
	 which contains the highest average pulsar flux in our collection.
	 {\bf Top panel}: 1961 periods containing 2000 highest peaks in the
	 file. {\bf Bottom panel}: The remaining 10\,164 in that file.
	 }
\label{fig4}
\end{figure}

It is further speculated here that the exact shape of the three principal 
components of the IP of \p0437 is a {\bf consequence} of the spiky emission, 
and not its {\bf cause}. This is supported by fig.~\ref{fig4}, where about 
16\% of the periods containing the 2000 strongest spikes in the entire data 
have an IP that has intense peaks in the three components; small variations 
in the spiky emission in any one of these three components is quite likely 
to change the shape of the overall IP of \p0437.

This further enhances the doubt, expressed in the last section
of Vivekanand \cite{MV2000}, about the interpretation of the
individual components of the IP of \p0437 as {\bf individual beams 
or regions of emission}. It is worth exploring the suggestion in that 
paper regarding the independent method of resolving the IP into 
independent beams or regions of emission on the pulsar.

\vfill
\eject
\vfill
\eject

\vfill
\eject

\vfill
\eject

\vfill
\eject

\end{document}